# Contribution of Data Categories to Readmission Prediction Accuracy

Wendong Ge, PhD[1, 3, 4], Hee Yeun Kim, PhD[1, 3], Sonali Desai, MD, MPH[1, 3],
Leonid Perlovsky, PhD[3, 4], Alexander Turchin, MD, MS[1, 2, 3]
[1]Brigham and Women's Hospital, Inc, Boston, MA
[2]Baim Institute for Clinical Research, Boston, MA
[3]Harvard Medical School, Boston, MA
[4]LP Information Technology Inc, Brookline, MA

**Abstract**

*Identification of patients at high risk for readmission could help reduce morbidity and mortality as well as healthcare costs. Most of the existing studies on readmission prediction did not compare the contribution of data categories. In this study we analyzed relative contribution of 90,101 variables across 398,884 admission records corresponding to 163,468 patients, including patient demographics, historical hospitalization information, discharge disposition, diagnoses, procedures, medications and laboratory test results. We established an interpretable readmission prediction model based on Logistic Regression in scikit-learn, and added the available variables to the model one by one in order to analyze the influences of individual data categories on readmission prediction accuracy. Diagnosis related groups (c-statistic increment of 0.0933) and discharge disposition (c-statistic increment of 0.0269) were the strongest contributors to model accuracy. Additionally, we also identified the top ten contributing variables in every data category.*

**Introduction**

Hospital readmission is an important metric of inpatient care quality and a major contributor to healthcare costs. Nearly 20% of hospitalized Medicare beneficiaries are readmitted to the hospital within 30 days of discharge[1]. For example, in 2009, the Medicare Payment Advisory Commission (MedPAC) found that Medicare spends about $12 billion per year on preventable readmissions[2]. In order to solve this problem, the Centers for Medicare and Medicaid Services (CMS) uses readmission rate as a criterion to penalize the hospitals whose readmission rate exceeds the expected threshold. In fiscal year 2017, more than half of the nation's hospitals (2597 hospitals) will be penalized for excessive readmission rates. The average penalty is 0.71% of the hospital's total Medicare reimbursement, and it can be as high as 3% depending on how far the rate of readmission exceeds the threshold. Additionally, readmission rate is also an important metric to evaluate the quality of healthcare. In the United States, readmission rate was selected as a significant indicator to measure the quality of healthcare in the National Quality Forum of 2008[3]. In the United Kingdom, readmission rate within 28 days after discharge is used to measure the quality of healthcare[4].

There are three categories of approaches to decreasing readmission rates: pre-discharge interventions, post-discharge interventions and bridging interventions[5]. Pre-discharge interventions may include patient education, discharge planning, medication reconciliation and follow-up appointment scheduling before discharge. Post-discharge interventions may include close follow-up, timely PCP communication, follow-up telephone call, patient hotline and home visits. Bridging interventions may include transition coach, patient-centered discharge instructions and provider continuity. However, it is very expensive to apply these interventions to every patient. Thus, it is significant to identify the patients at high risk for readmission in order to make these interventions cost-effective.

Electronic Medical Record (EMR) data provide abundant information to predict readmission risk. There are many published studies in this field. Hasan, Omar, et al.[6] performed logistic regression analysis to identify significant predictors of unplanned readmission within 30 days of discharge and developed a scoring system for estimating readmission risk. Shadmi, Efrat, et al[7]. developed a prediction score based on before admission electronic health record and administrative data using a preprocessing variable selection step with decision trees and neural network algorithms. Yu, Shipeng, et al.[8] proposed a generic framework for institution-specific readmission risk prediction, which takes patient data from a single institution and produces a statistical risk prediction model optimized for that particular institution and, optionally, for a specific condition. Chen, Robert, et al.[9] implemented a cloud-based predictive modeling system via a hybrid setup combining a secure private server with the Amazon Web Services Elastic Map-Reduce platform. Greenwald, Jeffrey L., et al.[10] designed a 30-day readmission risk prediction model through identification of physical, cognitive, and psychosocial issues using natural language processing. However, these investigations provide little information on the comparative importance of different data categories on the accuracy of the readmission prediction models. We therefore conducted a study to analyze the contributions of different data categories to the discriminative ability of readmission prediction.

**Methods**

**(1) Data Source**

We collected patient information for readmission prediction from Partners HealthCare, an integrated healthcare delivery system founded by Brigham and Women's Hospital and Massachusetts General Hospital, between 2000-01-01 and 2015-12-31. We obtained the EMR data from the Research Patient Data Registry (RPDR), a centralized clinical data registry or data warehouse that gathers clinical information from various Partners hospital systems.

**(2) The Definition of Readmission Prediction Problem**

Readmission prediction problem is defined as follows:

If a patient was hospitalized from admission day to discharge day, readmission prediction problem is to predict whether this patient would be admitted to hospital again in the next 30 days based on the features extracted from the EMR data up to the end of his index hospitalization, as illustrated in Figure 1.

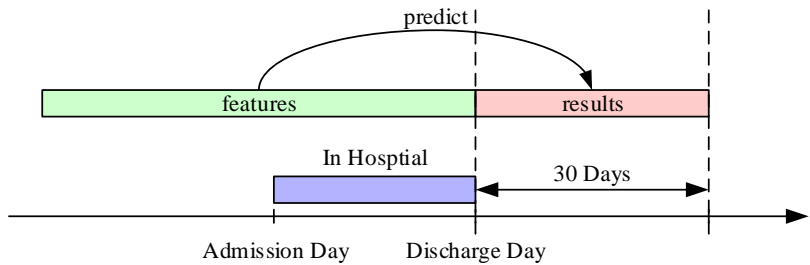

**Figure 1. Readmission prediction problem**

**(3) Readmission Records Generation**

In our dataset many patients had several admissions in his or her EMR data, which can be split into several readmission records for training or test. For example, Figure 2 is an admission history for a given patient. We can generate three readmission records for this patient:

- Record 1: use the features before 2010.10.20 to predict if he or she was readmitted to hospital from 2010.10.20 to 2010.11.19 (the result is No)

- Record 2: use the features before 2011.03.09 to predict if he or she was readmitted to hospital from 2011.03.09 to 2011.04.08 (the result is Yes)

- Record 3: use the features before 2011.03.25 to predict if he or she was readmitted to hospital from 2010.03.25 to 2010.04.24 (the result is No)

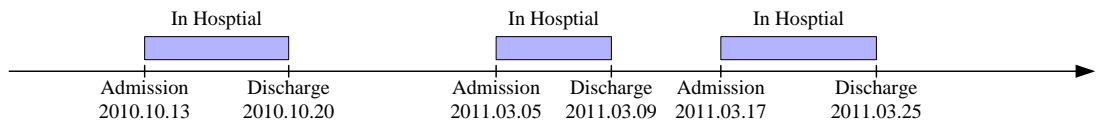

**Figure 2. Admission history**

**(4) Feature Description**

The features that we intended to analyze include patient demographics, historical hospitalization information, discharge disposition, diagnosis related group (DRG), diagnoses, procedures, medications and laboratory test results.

Demographics features include patient sex, age, race, and education level.

- Sex was represented as a binary feature (male / female)

- Age was represented as a categorical feature. We divided patient age into seven levels: 18-30, 31-40, 41-50, 51-60, 61-70, 71-80 and > 80.

- Race was represented as a categorical feature. The possible values were White, Black, Hispanic, Asian and Other.

- Education level was represented as a categorical feature. There were 16 possible values, including "DID NOT ATTEND SCHOOL", "8TH GRADE OR LESS", "HIGH SCHOOL GRADUATE/GED", "GRADUATED - COLLEGE", "GRADUATED - POST GRAD" and so forth.

Historical hospitalization information was parametrized as the number of hospitalizations in the year before the index admission.

Discharge Disposition is the patient's anticipated location or status after being discharged from the hospital[11]. In our data, discharge disposition was represented by a categorical feature with 45 possible values, including "Transferred to Chronic Hospital", "Transferred to Psych Hospital", "Dis/Trans to Medicaid-Certified Nursing Facility", "Dis/Trans to Home with IV Drug Therapy" and so forth.

DRG is a system of classifying the primary diagnoses for an inpatient stay into groups for the purposes of payment[12]. The DRG classification system divides possible diagnoses into more than 20 major body systems and subdivides them into almost 500 groups. Factors used to determine the DRG payment amount include the diagnosis involved as well as the hospital resources necessary to treat the condition. DRGs were represented by a categorical feature. Sometimes, a hospitalization is assigned more than one DRG code. We used unique combinations of DRG codes as possible values. Thus the number of possible values for the DRG feature in our data was 1657.

Diagnoses features were derived from the International Classification of Diseases version 9 (ICD-9) codes entered by clinicians treating the patient during the hospitalization[13]. International Classification of Diseases, as the short-form of International Statistical Classification of Diseases and Related Health Problems, is the international standard diagnostic ontology for epidemiology, health management and clinical purposes. This system is designed to map health conditions to corresponding generic categories together with specific variations, assigning for these a designated code, up to six characters long. In this work, we designed three diagnosis feature categories as follows

- ICD9-Prior-Binary: the existence of the target ICD-9 code at any time prior to admission.

- ICD9-Admission-Count: the number of times of the target ICD-9 code was recorded during the hospitalization.

- ICD9-Prior-Count: the number of times of the target ICD-9 code was recorded during the year prior to admission.

Procedures features were derived from the Current Procedural Terminology, 4th Edition (CPT-4) codes - a medical code set that is used to report medical, surgical, and diagnostic procedures and services to entities such as physicians, health insurance companies and accreditation organizations[14]. In this work, we designed two procedures feature categories as follows.

- CPT4-Prior-Binary: the existence of the target CPT-4 code at any time prior to admission.

- CPT4-Admission-Count: the number of times the target CPT-4 code was recorded during the hospitalization.

Medications features represented the information on medications administered during the hospitalization. In this work, each unique medication was represented as a binary feature, indicating whether the medication was given to the patient during the admission.

Laboratory test results were represented by four feature categories as follows.

- Labs-Admission-Binary: indicated whether the patient had the target lab test during the hospitalization.
- Labs-Prior-Binary: indicated whether the patient had the target lab test at any point prior to admission.
- Labs-Admission-Change: the slope of least-squares regression line through the target laboratory test results during the hospitalization. They are real features.
- Labs-Admission-SD: the standard deviation of the target lab test results during the hospitalization. They are non-negative real features.

The acquisition time frames for diagnoses, procedures, medications and laboratory test results features are illustrated in Figure 3.

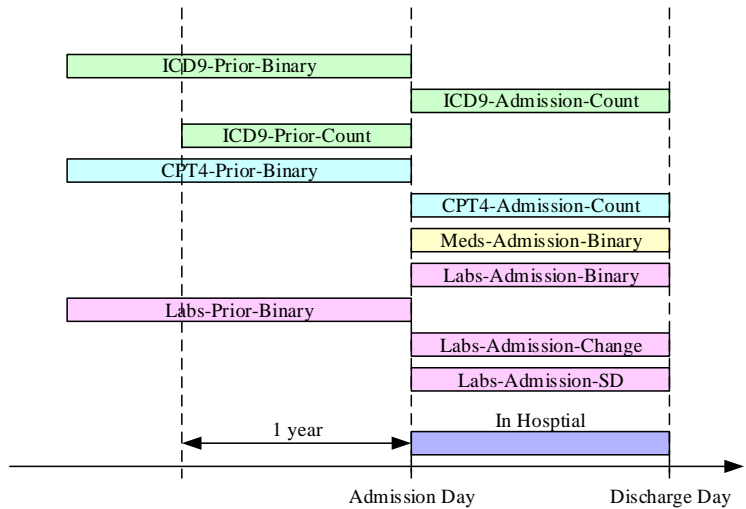

**Figure 3. Time frame of feature data acquisition**

**(5) Evaluation of Feature Category Contribution to Readmission Prediction Accuracy**

Logistic Regression in scikit-learn package was selected as our prediction model to provide a combination of clinical interpretability and an established machine learning tool[15]. One hot encoding (or Dummy variables) was used to transform categorical features into binary features. The number of binary features equals to the number of possible unique values of categorical features[16].

C-statistic was used to evaluate the discriminative ability of the readmission prediction model. As the first step, we generated the Receiver Operating Characteristic (ROC) curve by plotting the true positive rate (TPR) against the false positive rate (FPR) at various threshold settings. We subsequently calculated the c-statistic as the area under the ROC curve (AUC).

We tested the contribution to model accuracy of 17 feature categories, each of which was added to the model individually. Each iteration of the model was trained on the training data set (80% of records) and evaluated on the test data set (10% of records). Each feature category (with the exception of patient demographics; see below) was only retained in the subsequent iterations of the model if it resulted in a c-statistic improvement over the pre-defined threshold of 0.001.

**Results**

In this section, we discussed experimental results on contributions of different features to readmission prediction accuracy. The key points of the experiments are as follows:

- From the initial data set of 401,299 hospitalization records we excluded nine records without any diagnoses history prior to the hospitalization and 2,406 records of children under the age of 18. The remaining

- 398,884 hospitalization records corresponding to 163,468 patients were randomly divided (at the patient level) into training dataset (80% of patients), test dataset (10%) and validation dataset (10%).
- Each analytical record had information on the total of 90,101 features.
- The number of unique features in some of the data categories was very large. For example, there were 14885 ICD-9 codes in diagnoses features and 7887 CPT-4 codes in procedures features. If we used all the features to train the prediction model, the model would be over-fitting to the training dataset, decreasing the AUC for the test dataset. Thus, Pearson correlation coefficient was used to select the most predictive features in these data categories.
- The laboratory test results features had a large number of missing values. We therefore only included the features with no less than 10 records with values in our dataset, and imputed the missing values with median values across the dataset.

**(1) The Influence of Different Features on Readmission Risk Model Accuracy**

Table 1 illustrates the contributions of different feature categories to readmission prediction model accuracy. The final model reached c-statistic of over 0.7. The summary of the results are as follows:

- "Sex", "Age" and "Race" are basic demographics features. They were kept in the model regardless of the associated changes in model accuracy. The c-statistic based on these three features alone is 0.5413.
- "Educational Level" is a predictive feature. Its improvement is 0.009
- "Hospitalization History" was not an effectively predictive feature and its improvement was only 0.0002.
- "Discharge Disposition" and "DRG" were effectively predictive features, and the c-statistic improvements corresponding to them were 0.0269 and 0.0933 respectively.
- Among the diagnoses features, "ICD9-Prior-Binary" was a more predictive feature, compared with "ICD9-Prior-Count" and "ICD9-Prior-Count". It was associated with a c-statistic increase of 0.0158.
- Among the procedures features, "CPT4-Admission-Count" was more important than "CPT4-Prior-Binary" and was associated with an increase in c-statistic of 0.0063.
- "Meds-Admission-Binary" was an effectively predictive feature. It was associated with a c-statistic increase of 0.0062
- Among the labs features, the existence of lab tests before admission (+0.0022) was more important than the existence of lab test during admission (-0.0005). However, during hospitalization, the changes (+0.0019) and standard deviations (+0.0012) of the laboratory test results were predictive factors.

**Table 1. Contributions of feature categories to readmission prediction accuracy**

| Feature category | type | number | Selected (Pearson) | AUC | Improvement (Th = 0.001) | action |
|---|---|---|---|---|---|---|
| Sex | binary | 1 | NA | 0.5231 | NA | keep |
| Age | categorical | 7 | NA | 0.5423 | 0.0192 | keep |
| Race | categorical | 5 | NA | 0.5413 | -0.0010 | keep |
| Education Level | categorical | 16 | NA | 0.5503 | 0.0090 | keep |
| Hospitalization History | quantitative | 1 | NA | 0.5505 | 0.0002 | remove |
| Discharge Disposition | categorical | 45 | NA | 0.5772 | 0.0269 | keep |
| DRG | categorical | 1657 | NA | 0.6705 | 0.0933 | keep |
| ICD9-Prior-Binary | binary | 14885 | 100 | 0.6863 | 0.0158 | keep |
| ICD9-Admission-Count | quantitative | 14885 | 20 | 0.6871 | 0.0008 | remove |
| ICD9-Prior-Count | quantitative | 14885 | 200 | 0.6858 | -0.0005 | remove |
| CPT4-Prior-Binary | binary | 7887 | 50 | 0.6871 | 0.0008 | remove |
| CPT4-Admission-Count | quantitative | 7887 | 50 | 0.6926 | 0.0063 | keep |
| Meds-Admission-Binary | binary | 10618 | 250 | 0.6988 | 0.0062 | keep |
| Labs-Admission-Binary | binary | 5774 | 10 | 0.6983 | -0.0005 | remove |
| Labs-Prior-Binary | binary | 5774 | 30 | 0.7010 | 0.0022 | keep |

| Labs-admission-Change | real | 5774 | 4 | 0.7029 | 0.0019 | keep |
| Labs-admission-SD | non-negative real | 5774 | 4 | 0.7041 | 0.0012 | keep |

**(2) Contributions of Individual Features**

Table 2 illustrates top ten feature-value pairs among all feature categories by contribution to the model. We used the corresponding coefficients in the Logistic Regression model as the indicators of the importance of the features to the model of readmission risk. Among these top ten features, seven features were DRGs and three features were Discharge Disposition values.

**Table 2. Top ten contributing feature-value pairs**

| No. | Category | Feature name | Feature meaning | | coefficient |
|---|---|---|---|---|---|
| 1 | DRG | DRG: 379 | MED | gastrointestinal hemorrhage w/o CC/MCC | 2.0893 |
| 2 | DRG | DRG: 384 | MED | Uncomplicated peptic ulcer w/o MCC | 1.7685 |
| 3 | DRG | DRG: 383 | MED | Uncomplicated peptic ulcer w MCC | 1.6675 |
| 4 | Discharge Disposition | Discharge to Psychiatric Hospital | NA | | 1.6009 |
| 5 | DRG | DRG: 172 | Med | Digestive malignancy w MCC | 1.2470 |
| 6 | DRG | DRG: 203 | MED | Bronchitis & asthma w/o CC/MCC | 1.1171 |
| 7 | DRG | DRG: 410 | SURG | Biliary tract proc except only cholecyst w or w/o c.d.e. w/o CC/MCC | 0.8683 |
| 8 | Discharge Disposition | Hospice Medical Facility | NA | | 0.8466 |
| 9 | DRG | DRG: 010 | SURG | Pancreas transplant | 0.7825 |
| 10 | Discharge Disposition | Dis/Trans to Short-Term Hospital as IP (inpatient) | NA | | 0.7214 |

Figures 4 and 5 illustrate the contribution of different education levels and age categories to readmission prediction accuracy, respectively.

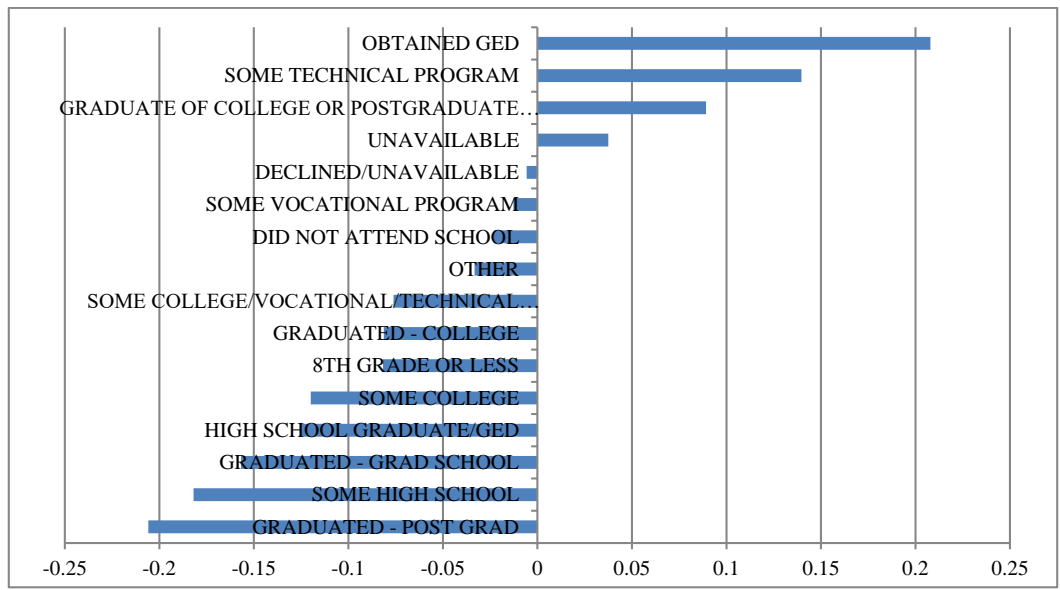

**Figure 4. Contributions of different education levels to readmission prediction accuracy**

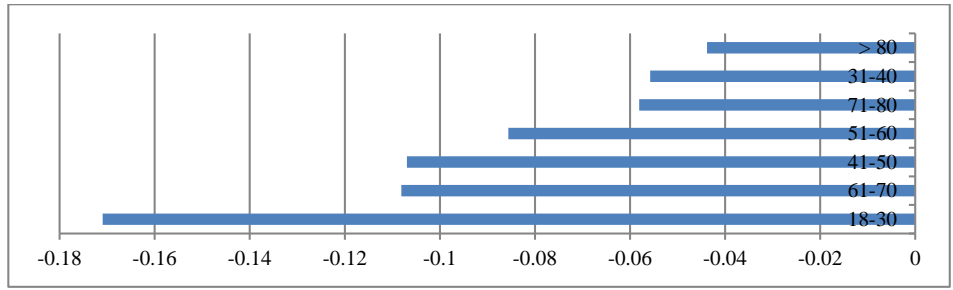

**Figure 5. Contributions of different age categories readmission prediction accuracy**

Tables 3 through 10 describe the top ten features / feature-value pairs by contribution to the model accuracy in the following categories: discharge disposition, DRG, ICD9-Prior-Binary, CPT4-Admission-Count, Meds-Admission-Binary, Labs-Prior-Binary, Labs-Admission-Change and Labs-Admission-SD respectively.

**Table 3. Top ten contributing Discharge Disposition values**

| No. | Value | coefficient |
|---|---|---|
| 1 | Discharge to Psychiatric Hosp | 1.6009 |
| 2 | Hospice Medical Facility | 0.8466 |
| 3 | Dis/Trans to Short-Term Hospital as IP | 0.7214 |
| 4 | Short Term General Hospital | 0.4811 |
| 5 | Dis/Trans to Long Term Care Hospital | 0.2917 |
| 6 | Against Medical Advice | 0.2879 |
| 7 | Dis/Trans to Skilled Nursing Facility w/Approval | 0.2876 |
| 8 | Dis/Trans to Inpatient Rehab Facility | 0.2371 |
| 9 | Transferred to Level 1 | 0.1281 |
| 10 | Left Against Medical Advice or Patient Discontinued Care | 0.1264 |

**Table 4. Top ten contributing DRG values**

| No. | DRG code | | DRG code description | coefficient |
|---|---|---|---|---|
| 1 | 379 | MED | G.I. hemorrhage w/o CC/MCC | 2.0893 |
| 2 | 384 | MED | Uncomplicated peptic ulcer w/o MCC | 1.7685 |
| 3 | 383 | MED | Uncomplicated peptic ulcer w MCC | 1.6675 |
| 4 | 172 | Med | Digestive malignancy w MCC | 1.2470 |
| 5 | 203 | MED | Bronchitis & asthma w/o CC/MCC | 1.1171 |
| 6 | 410 | SURG | Biliary tract proc except only cholecyst w or w/o c.d.e. w/o CC/MCC | 0.8683 |
| 7 | 010 | SURG | Pancreas transplant | 0.7825 |
| 8 | 082 | MED | Traumatic stupor & coma, coma >1 hr w MCC | 0.7072 |
| 9 | 366 | MED | Malignancy, female reproductive system w MCC | 0.6292 |
| 10 | 403 | MED | Lymphoma & non-acute leukemia w MCC | 0.5727 |

**Table 5. Top ten contributing ICD9-Prior-Binary features**

| No. | Feature name | Feature meaning | coefficient |
|---|---|---|---|
| 1 | 654.50 | Cervical incompetence, unspecified as to episode of care or not applicable | 0.5671 |
| 2 | 155.1 | Malignant neoplasm of intrahepatic bile ducts | 0.4260 |
| 3 | 372.31 | Rosacea conjunctivitis | 0.2631 |
| 4 | 511.81 | Malignant pleural effusn | 0.2563 |
| 5 | 864.15 | Unspecified laceration of liver with open wound into cavity | 0.2408 |
| 6 | 305.42 | Sedative, hypnotic or anxiolytic abuse, episodic | 0.2320 |
| 7 | 864.14 | Liver lacerat, major-opn | 0.2313 |
| 8 | 789.51 | Malignant ascites | 0.1783 |

| 9 | 162.9 | Malignant neoplasm of bronchus and lung, unspecified | 0.1748 |
|---|---|---|---|
| 10 | 651.50 | Quads w fetal loss-unsp | 0.1737 |

**Table 6. Top ten contributing CPT4-Admission-Count features**

| No. | Feature name | Feature meaning | coefficient |
|---|---|---|---|
| 1 | 77418 | Intensity modulated treatment delivery, single or multiple fields/arcs, via narrow spatially and temporally modulated beams, binary, dynamic MLC, per treatment session | 0.5451 |
| 2 | 59051 | Fetal monitoring during labor by consulting physician (ie, non-attending physician) with written report; interpretation only | 0.2774 |
| 3 | 77336 | Continuing medical physics consultation, including assessment of treatment parameters, quality assurance of dose delivery, and review of patient treatment documentation in support of the radiation oncologist, reported per week of therapy | 0.2350 |
| 4 | 59510 | Routine obstetric care including antepartum care, cesarean delivery, and postpartum care | 0.2046 |
| 5 | 77427 | Radiation treatment management, five treatments | 0.1335 |
| 6 | 76816 | Ultrasound, pregnant uterus, real time with image documentation, follow-up (eg, re-evaluation of fetal size by measuring standard growth parameters and amniotic fluid volume, re-evaluation of organ system(s) suspected or confirmed to be abnormal on a previous scan), transabdominal approach, per fetus | 0.1070 |
| 7 | 01967 | Neuraxial labor analgesia/anesthesia for planned vaginal delivery (this includes any repeat subarachnoid needle placement and drug injection and/or any necessary replacement of an epidural catheter during labor) | 0.0983 |
| 8 | 99254 | Inpatient consultation for a new or established patient, which requires these three key components: A comprehensive history; A comprehensive examination; and Medical decision making of moderate complexity. Counseling and/or coordination of care with other providers or agencies are provided consistent with the nature of the problem(s) and the patient's and/or family's needs. Usually, the presenting problem(s) are of moderate to high severity. Physicians typically spend 80 minutes at the bedside and on the patient's hospital floor or unit. | 0.0772 |
| 9 | 99255 | Inpatient consultation for a new or established patient, which requires these three key components: A comprehensive history; A comprehensive examination; and Medical decision making of high complexity. Counseling and/or coordination of care with other providers or agencies are provided consistent with the nature of the problem(s) and the patient's and/or family's needs. Usually, the presenting problem(s) are of moderate to high severity. Physicians typically spend 110 minutes at the bedside and on the patient's hospital floor or unit. | 0.0638 |
| 10 | 87324 | Infectious agent antigen detection by immunoassay technique, (eg, enzyme immunoassay [EIA], enzyme-linked immunosorbent assay [ELISA], immunochemiluminometric assay [IMCA]) qualitative or semiquantitative, multiple-step method; Clostridium difficile toxin(s) | 0.0629 |

**Table 7. Top ten contributing Meds-Admission-Binary features**

| No. | Feature name | Feature meaning | coefficient |
|---|---|---|---|
| 1 | MGH_CC-770900 | Haloperidol 5mg tablet | 0.4631 |
| 2 | MGH_CC-815200 | Lido hcl 0 < 5% w/ma | 0.4513 |
| 3 | MGH_CC-000524 | Hydromor 20 mcg/ml bup 0.1% epd | 0.3880 |
| 4 | MGH_CC-713540 | Blenoxane 15 u powder for injection | 0.3869 |
| 5 | MGH_CC-710060 | Bacitracin ophthalmic 500 u/gm ointment | 0.3650 |
| 6 | MGH_CC-000189 | Ca leukovorin 50 mg | 0.3613 |
| 7 | BWH_CC-947765 | Hydromorphone 50mg in 0.9% sodium chloride 250ml bag | 0.3503 |

| 8 | MGH_CC-610700 | Glass bottle 500 ml | 0.3447 |
| 9 | MGH_CC-000039 | Carboplatin 50 mg | 0.3268 |
| 10 | MGH_CC-823020 | Methotrexate sodium 25 mg/ml solution | 0.3260 |

**Table 8. Top ten contributing Labs-Prior-Binary features**

| No. | Feature name | Feature meaning | coefficient |
|---|---|---|---|
| 1 | QMOTIL | Motility quality, sperm | 0.3173 |
| 2 | NRBC% | NRBC(%) | 0.1348 |
| 3 | CardRR | Relative CHD Risk (LDL/HDL) | 0.1098 |
| 4 | BARB-TS | Barbiturates (Tox Screen) | 0.0882 |
| 5 | FE | Iron level | 0.0802 |
| 6 | GFR | Glomerula Filtration Rate (estimated) | 0.0651 |
| 7 | CASTS | Casts (/HPF) in urine | 0.0482 |
| 8 | XCHROM | X Chromosome Gene analysis | 0.0478 |
| 9 | TB-GENO | Mycobacteria tuberculosis genotype | 0.0475 |
| 10 | ALPINT | Interpretation (ALPS) | 0.0473 |

**Table 9. Top contributing Labs-Admission-Change features**

| No. | Test code | Test description | coefficient |
|---|---|---|---|
| 1 | RBC | Red blood cell count | 0.1069 |
| 2 | HCT | Hematocrit | -0.0024 |
| 3 | CD8AB | Absolute CD8+ count | -0.0038 |
| 4 | DIAZ | Diazepam level | -0.2597 |

**Table 10. Top contributing Labs-Admission-SD features**

| No. | Feature name | Feature meaning | coefficient |
|---|---|---|---|
| 1 | RDW | Red blood cell distribution width | 0.0330 |
| 2 | PLT | Platelet count | -4.969E-05 |
| 3 | POLYS | Neutrophil count | -0.0141 |
| 4 | 3+PCT | CD3% | -0.9308 |

**Discussion**

In this paper, we established a prediction model for the patients at high risk for readmission and analyzed the effects of different feature categories, including sex, age, race, education level, historical hospitalization information, discharge disposition, diagnosis related group (DRG), diagnoses, procedures, medications and laboratory test results. From the experimental results, we found out the following conclusions:

- Demographic characteristics of patients are not a significant contributor to readmission prediction

- DRG and discharge disposition are the most important features to readmission prediction.

- The diagnoses preceding hospitalization (represented by ICD-9 codes prior to admission) were more important than the ICD-9 codes recorded during the admission and a year before the admission in readmission prediction task.

- The information about procedures performed during the hospitalization (represented by CPT-4 codes) was more important than the information about procedures performed prior to admission.

- Change and variation (represented by standard deviation) of laboratory test results were more important than the mere existence of laboratory tests during hospitalization.

Our findings have to be interpreted in the light of several limitations. Our data was limited to a single healthcare system, and thus may not be generalizable to other institutions / settings. Some of our data (e.g. discharge disposition / DRG) would only become available after discharge and therefore its utility for identifying patients at high risk for readmission while still hospitalized may be limited. The discriminatory ability of the resulting model, while in line with what has been described in other studies, may not be sufficient for practical applications.

In conclusion, we have analyzed contribution of multiple data categories to identification of patients at high risk for hospital readmission, including several (e.g. laboratory test result change / variation during hospitalization) that have not been widely investigated before. We have identified several data categories whose contribution was particularly important. Further research is needed to continue to identify information categories critical for detection of patients at high risk for hospital readmissions.

**Acknowledgements**

This work was supported in part by grants from Brigham and Women's Hospital Department of Medicine Innovation Evergreen Fund, Agency for Healthcare Research and Quality R01HS024090 and NIDDK R41 DK105612.

**References**

1. Joynt KE, Orav EJ, Jha AK. Thirty-day readmission rates for Medicare beneficiaries by race and site of care. JAMA. 2011 Feb 16;305(7):675-81.
2. Monette M. Hospital readmission rates under the microscope. CMAJ. 2012 Sep 04;184(12):E651-2.
3. Ross JS, Chen J, Lin Z, Bueno H, Curtis JP, Keenan PS, et al. Recent national trends in readmission rates after heart failure hospitalization. Circ Heart Fail. 2010 Jan;3(1):97-103.
4. Lyratzopoulos G, Havely D, Gemmell I, Cook GA. Factors influencing emergency medical readmission risk in a UK district general hospital: a prospective study. BMC Emerg Med. 2005 Jul 18;5:4.
5. Tang N, Fujimoto J, Karliner L. Evaluation of a primary care-based post-discharge phone call program: keeping the primary care practice at the center of post-hospitalization care transition. J Gen Intern Med. 2014 Nov;29(11):1513-8.
6. Hasan O, Meltzer DO, Shaykevich SA, Bell CM, Kaboli PJ, Auerbach AD, et al. Hospital readmission in general medicine patients: a prediction model. J Gen Intern Med. 2010 Mar;25(3):211-9.
7. Shadmi E, Flaks-Manov N, Hoshen M, Goldman O, Bitterman H, Balicer RD. Predicting 30-day readmissions with preadmission electronic health record data. Med Care. 2015 Mar;53(3):283-9.
8. Yu S, Farooq F, van Esbroeck A, Fung G, Anand V, Krishnapuram B. Predicting readmission risk with institution-specific prediction models. Artif Intell Med. 2015 Oct;65(2):89-96.
9. Chen R, Su H, Khalilia M, Lin S, Peng Y, Davis T, et al. Cloud-based Predictive Modeling System and its Application to Asthma Readmission Prediction. AMIA Annu Symp Proc. 2015;2015:406-15.
10. Greenwald JL, Cronin PR, Carballo V, Danaei G, Choy G. A Novel Model for Predicting Rehospitalization Risk Incorporating Physical Function, Cognitive Status, and Psychosocial Support Using Natural Language Processing. Med Care. 2017 Mar;55(3):261-6.
11. Bini SA, Fithian DC, Paxton LW, Khatod MX, Inacio MC, Namba RS. Does discharge disposition after primary total joint arthroplasty affect readmission rates? J Arthroplasty. 2010 Jan;25(1):114-7.
12. Chuang KH, Covinsky KE, Sands LP, Fortinsky RH, Palmer RM, Landefeld CS. Diagnosis-related group-adjusted hospital costs are higher in older medical patients with lower functional status. J Am Geriatr Soc. 2003 Dec;51(12):1729-34.
13. Rosen LM, Liu T, Merchant RC. Efficiency of International Classification of Diseases, Ninth Revision, billing code searches to identify emergency department visits for blood or body fluid exposures through a statewide multicenter database. Infect Control Hosp Epidemiol. 2012 Jun;33(6):581-8.
14. Bentley PN, Wilson AG, Derwin ME, Scodellaro R, Jackson RE. Reliability of assigning correct current procedural terminology-4 E/M codes. Ann Emerg Med. 2002 Sep;40(3):269-74.
15. Abraham A, Pedregosa F, Eickenberg M, Gervais P, Mueller A, Kossaifi J, et al. Machine learning for neuroimaging with scikit-learn. Front Neuroinform. 2014;8:14.
16. Arroyo Lopez FN, Duran Quintana MC, Garrido Fernandez A. Use of logistic regression with dummy variables for modeling the growth-no growth limits of Saccharomyces cerevisiae IGAL01 as a function of sodium chloride, acid type, and potassium sorbate concentration according to growth media. J Food Prot. 2007 Feb;70(2):456-65.